\documentclass[aps,prl,preprint,tightenlines,superscriptaddress,showpacs,byrevtex]{revtex4}
\usepackage{graphicx}
\usepackage{dcolumn}
\usepackage{bm}

\begin{document}

\vspace*{-3\baselineskip}
\resizebox{!}{3cm}{\includegraphics{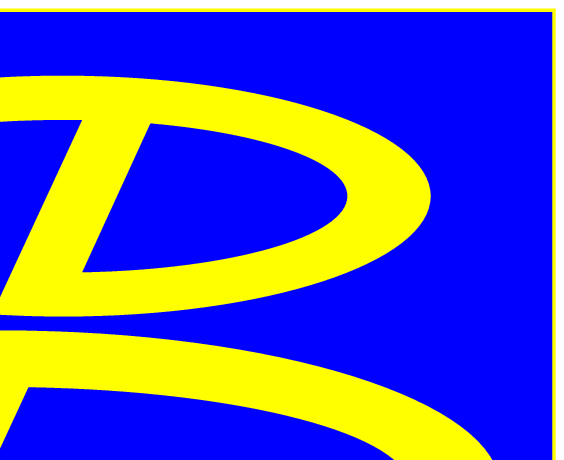}}

\preprint{KEK Preprint 2002-128}
\preprint{Belle Preprint 2002-40}

\def\tbzresultmean{1.554}
\def\tbzresultstat{\pm 0.030}
\def\tbzresultsyst{\pm0.019}

\def\tbmresultmean{1.695}
\def\tbmresultstat{\pm 0.026}
\def\tbmresultsyst{\pm 0.015}

\def\rbmresultmean{1.091}
\def\rbmresultstat{\pm 0.023}
\def\rbmresultsyst{\pm 0.014}

\def\yresultmean{0.114}
\def\yresultstat{{_{-0.064}^{+0.060}}}
\def\yresultsyst{{_{-0.011}^{+0.012}}}

\def\yinterval{-0.013 < y_{CP} < 0.234}

\def\tbzresult{\tbzresultmean \tbzresultstat\;\mbox{(stat)} \; \tbzresultsyst\;\mbox{(syst)}}
\def\tbmresult{\tbmresultmean \tbmresultstat\;\mbox{(stat)} \tbmresultsyst\;\mbox{(syst)}}
\def\rbmresult{\rbmresultmean \rbmresultstat\;\mbox{(stat)} \rbmresultsyst\;\mbox{(syst)}}
\def\yresult{\yresultmean \; \yresultstat \; \yresultsyst}


\def\rtau{r_\tau}

\def\yllimit{-0.36}
\def\yulimit{0.35}
\def\tbzdstlnu{1.50\pm0.06^{+0.06}_{-0.04}}
\def\tbmdstlnu{1.54\pm0.10^{+0.14}_{-0.07}}
\def\tbzdstpm{1.55^{+0.18}_{-0.17}{}^{+0.10}_{-0.07}}
\def\tbzdppm{1.41^{+0.13}_{-0.12}\pm0.07}
\def\tbmdzpm{1.73\pm0.10\pm0.09}
\def\tbmpsikm{1.87^{+0.13}_{-0.12}{}^{+0.07}_{-0.14}}
\def\tbzpsiks{1.54^{+0.28}_{-0.24}{}^{+0.11}_{-0.19}}
\def\tbzpsikst{1.56^{+0.22}_{-0.19}{}^{+0.09}_{-0.15}}
\def\Tbzdstlnu{$\taubz=(\tbzdstlnu)$~ps}
\def\Tbmdstlnu{$\taubm=(\tbmdstlnu)$~ps}
\def\Tbzdstpm{$\taubz=(\tbzdstpm)$~ps}
\def\Tbzdppm{$\taubz=(\tbzdppm)$~ps}
\def\Tbmdzpm{$\taubm=(\tbmdzpm)$~ps}
\def\Tbmpsikm{$\taubm=(\tbmpsikm)$~ps}
\def\Tbzpsiks{$\taubz=(\tbzpsiks)$~ps}
\def\Tbzpsikst{$\taubz=(\tbzpsikst)$~ps}
\def\Tbzresult{$\taubz=(\tbzresult)$~ps}
\def\Tbmresult{$\taubm=(\tbmresult)$~ps}
\def\Rbmresult{$\rbm=\rbmresult$}
\def\Yresult{$\ycp=\yresult$}
\def\Ylimit{$\yllimit<\ycp<\yulimit$}
\def\intl{5.1}


\newcommand{\nim}[3]{Nucl. Inst. and Meth. {\bf #1} #2 (#3)}
\newcommand{\prld}[3]{Phys. Rev. Lett. {\bf #1} #2 (#3)}
\newcommand{\prdd}[3]{Phys. Rev. D {\bf #1} #2 (#3)}
\newcommand{\prxd}[3]{Phys. Rev. {\bf #1} #2 (#3)}
\newcommand{\plb}[3]{Phys. Lett. B {\bf #1} #2 (#3)}
\newcommand{\npa}[3]{Nucl. Phys. A {\bf #1} #2 (#3)}
\newcommand{\apphyslet}[3]{App. Phys. Lett. {\bf #1} #2 (#3)}
\newcommand{\zphysa}[3]{Z. Phys. A. {\bf #1} #2 (#3)}
\newcommand{\zpc}[3]{Z. Phys. C {\bf #1} #2 (#3)}
\newcommand{\jphys}[3]{J. Phys. {\bf #1} #2 (#3)}

\def\etal{{\it et al.}}
\def\chisqndf{{\chi}^2/n}
\def\chisq{\chi^2}
\def\Chisqndf{$\chisqndf$}
\def\Chisq{$\chisq$}
\def\NDF{$N.D.F.$}
\def\degree{{}^{\circ}}
\def\dG{\Delta\Gamma}
\def\dM{\Delta M}
\def\rmix{R_{\rm mix}}
\def\amix{A_{\rm mix}}
\def\re{{\cal R}e}
\def\im{{\cal I}m}
\def\bzb{{\overline{B}{}^0}}
\def\bb{{\overline{B}{}^0}}
\def\kz{{K{}^0}}
\def\kzb{{\overline{K}{}^0}}
\def\kb{{\overline{K}{}^0}}
\def\bz{{B^0}}
\def\bh{{B_H}}
\def\bl{{B_L}}
\def\bm{{B^-}}
\def\bp{{B^+}}
\def\taubz{\tau(\bzb)}
\def\taubm{\tau(\bm)}
\def\rbm{\taubm/\taubz}
\def\Bzb{$\bzb$}
\def\Bm{$\bm$}
\def\Fb{fb$^{-1}$}
\def\ycp{y_{CP}}
\def\piz{\pi^0}
\def\pip{\pi^+}
\def\pim{\pi^-}
\def\kz{K^0}
\def\kp{K^+}
\def\km{K^-}
\def\ks{K_S^0}
\def\kl{K_L^0}
\def\kb{\overline{K}}
\def\rhom{\rho^-}
\def\bbar{\overline{B}}
\def\bbbar{B\bbar}
\def\bzbzbar{\bz\bzb}
\def\BBbar{$\bbbar$}
\def\BzBzbar{$\bzbzbar$}
\def\ccbar{c\overline{c}}
\def\CCbar{$\ccbar$}
\def\dstar{D^{*}}
\def\dstarz{{D^{*0}}}
\def\dstarp{D^{*+}}
\def\dstarzb{\overline{D}^{*0}}
\def\dstarm{D^{*-}}
\def\dmdstp{\Delta M_{\dstarp}}
\def\dmdstz{\Delta M_{\dstarz}}
\def\dmdst{\Delta M_{\dstar}}
\def\DM{$\Delta M$}
\def\Dstar{$\dstar$}
\def\Dstarz{$\dstarz$}
\def\Dstarp{$\dstarp$}
\def\Dstarzb{$\dstarzb$}
\def\Dstarm{$\dstarm$}
\def\nub{\overline{\nu}}
\def\jpsi{{J/\psi}}
\def\dz{D^0}
\def\Dz{$\dz$}
\def\Dt{\Delta t}
\def\Dz{\Delta z}
\def\dplus{D^+}
\def\Dp{$\dplus$}
\def\dzb{\overline{D}{}^0}
\def\Dzb{$\dzb$}
\def\dstl{\dstar\ell}
\def\dstpl{\dstarp\ell}
\def\dstzl{\dstarz\ell}
\def\dstml{\dstarm\ell}
\def\dstzbl{\dstarzb\ell}
\def\bdstlnu{\overline{B}\to\dstar\ell^-\nub}
\def\bdstxlnu{\overline{B}\to\dstar X\ell^-\nub}
\def\bzdstlnu{\bzb\to\dstarp\ell^-\nub}
\def\bzdstxlnu{\bzb\to\dstarp X\ell^-\nub}
\def\bmdstlnu{\bm\to\dstarz\ell^-\nub}
\def\bmdstxlnu{\bm\to\dstarz X\ell^-\nub}
\def\bmdstpxlnu{\bm\to\dstarp X\ell^-\nub}
\def\bpsik{\overline{B}\to\jpsi\kb}
\def\bdpi{\bbar\to D\pi}
\def\bzdstpm{\bzb\to\dstarp\pim}
\def\bzdstrhom{\bzb\to\dstarp\rhom}
\def\bzdppm{\bzb\to\dplus\pim}
\def\bmdzpm{\bm\to\dz\pim}
\def\bpsiks{B\to\jpsi\ks}
\def\bzbpsiks{\bzb\to\jpsi\ks}
\def\bzpsiks{\bz\to\jpsi\ks}
\def\bzbpsikst{\bzb\to\jpsi\kstarzb}
\def\bmpsikm{\bm\to\jpsi\km}
\def\nbdstlnu{N_{\bdstlnu}}
\def\nbdstxlnu{N_{\bdstxlnu}}
\def\nbb{N_{\bbbar}}
\def\ncc{N_{\ccbar}}
\def\kpi{K^-\pi^+}
\def\kpipiz{\km\pip\piz}
\def\kpipipi{\km\pip\pip\pim}
\def\kstarzb{\overline{K}{}^{*0}}
\def\Kstarzb{$\kstarzb$}
\def\UPS{$\Upsilon(4S)$}
\def\bgu{(\beta\gamma)_\Upsilon}
\def\pdstl{\mathbf{p}_{\dstar\ell}}
\def\pb{\mathbf{p}_{B}}
\def\Gevc{GeV/$c$}
\def\Gevcsq{GeV/$c^2$}
\def\Mevc{MeV/$c$}
\def\Mevcsq{MeV/$c^2$}
\def\degree{{}^{\rm o}}
\def\Degree{${}^{\rm o}$}
\def\dE{{\Delta E}}
\def\DE{$\dE$}
\def\mb{{M_{\rm bc}}}
\def\Mb{$\mb$}
\def\micron{$\mu$m}
\def\delz{\Delta z}
\def\delt{\Delta z}
\def\dt{\Delta t}
\def\dzb{\delz_B}
\def\dtb{\dt_B}
\def\dtp{\dt'}
\def\DTb{$\dtb$}
\def\dtrec{\dt_{rec}}
\def\dtgen{\dt_{gen}}
\def\dzrec{\delz_{rec}}
\def\dzgen{\delz_{gen}}
\def\sigdt{\sigma_{\dt}}
\def\sigpdt{\sigma'_{\dt}}
\def\sigmisdt{\sigma_{tail}^{\dt}}
\def\sigdz{\sigma_{\delz}}
\def\sigtdz{\tilde{\sigma}_{\delz}}
\def\sigtmisdz{\tilde{\sigma}_{tail}^{\delz}}
\def\arec{\alpha_{rec}}
\def\aasc{\alpha_{asc}}
\def\sigzrec{\sigma_{z}^{rec}}
\def\sigzasc{\sigma_{z}^{asc}}
\def\sigtzrec{\tilde{\sigma}_{z}^{rec}}
\def\sigtzasc{\tilde{\sigma}_{z}^{asc}}
\def\sigk{\sigma_{K}}
\def\sigmisk{\sigma_{tail}^{K}}
\def\smis{S_{tail}}
\def\scharm{S_{charm}}
\def\sdet{S_{det}}
\def\sdata{s_{\rm data}}
\def\scmis{S_{tail}^{charm}}
\def\sdmis{S_{tail}^{det}}
\def\smisbg{S_{tail}^{\rm bkg}}
\def\sbg{S_{\rm bkg}}
\def\mudz{\mu_{\delz}}
\def\mumisdz{\mu_{tail}^{\delz}}
\def\mumisbg{\mu_{tail}^{\rm bkg}}
\def\muz{\mu_0}
\def\mumisz{\mu_{tail}^0}
\def\mudt{\mu_{\dt}}
\def\mumisdt{\mu_{tail}^{\dt}}
\def\amu{\alpha_\mu}
\def\amismu{\alpha_{tail}^{\mu}}
\def\mubg{\mu_{\rm bkg}}
\def\fmis{f_{tail}}
\def\fmisbg{f_{tail}^{\rm bkg}}
\def\flmbg{f_{\lambda \rm bkg}}
\def\fsig{f_{\rm sig}}
\def\fbkg{f_{\rm bkg}}
\def\fbg{f_{\rm bkg}}
\def\Fbg{F_{\rm bkg}}
\def\Fsig{F_{\rm sig}}
\def\lmbg{\lambda_{\rm bkg}}
\def\pbg{p_{\rm bkg}}
\def\fm{f_-}
\def\fz{f_0}
\def\ffdst{f_{f \dstar}}
\def\pfdst{p_{f \dstar}}
\def\lfdst{\lambda_{f \dstar}}
\def\ffl{f_{f \ell}}
\def\fbb{f_{\bbbar}}
\def\fcc{f_{\ccbar}}
\def\ftbg{f_{\tau {\rm bkg}}}
\def\frdstl{f_{R(\dstar\ell)}}
\def\fbgde{F_{\rm bkg}^{\dE}}
\def\fsigde{F_{\rm sig}^{\dE}}
\def\fbgmb{F_{\rm bkg}^{\mb}}
\def\fsigmb{F_{\rm sig}^{\mb}}
\def\rsig{R_{\rm sig}}
\def\rfdst{R_{f \dstar}}
\def\pfl{p_{f \ell}}
\def\pbb{p_{\bbbar}}
\def\pcc{p_{c\overline{c}}}
\def\Rsig{$\rsig$}
\def\rdz{R_{\delz}}
\def\rbg{R_{\rm bkg}}
\def\Rdz{$\rdz$}
\def\tbm{\tau_-}
\def\tbz{\tau_0}
\def\tsig{\tau_{\rm sig}}
\def\tbg{\tau_{\rm bkg}}
\def\BF{{\cal B}}
\def\cosb{\cos\theta_B}
\def\ie{{\it i.e.}}
\def\psig{{{\cal P}_{\rm sig}}}
\def\pbkg{{{\cal P}_{\rm bkg}}}
\def\pasym{{{\cal P}_{\rm asym}}}
\def\rdet{{{R}_{\rm det}}}
\def\rnp{{{R}_{\rm np}}}
\def\rk{{{R}_{\rm k}}}

\newcommand{\Bcp}{B_{CP}}
\newcommand{\Btag}{B_{\rm tag}}
\newcommand{\zcp}{z_{CP}}
\newcommand{\ztag}{z_{\rm tag}}

\newcommand{\Brec}{B_{\rm rec}}
\newcommand{\Basc}{B_{\rm asc}}
\newcommand{\trec}{t_{\rm rec}}
\newcommand{\tasc}{t_{\rm asc}}
\newcommand{\zrec}{z_{\rm rec}}
\newcommand{\zasc}{z_{\rm asc}}
\newcommand{\Rrec}{R_{\rm rec}}
\newcommand{\Rasc}{R_{\rm asc}}
\newcommand{\Srec}{\sigma_{\rm rec}}
\newcommand{\Sasc}{\sigma_{\rm asc}}
\newcommand{\Scp}{\sigma_{CP}}
\newcommand{\Stag}{\sigma_{\rm tag}}
\newcommand{\Sdz}{\sigma_{\Dz}}
\newcommand{\Sdt}{\sigma_{\Dt}}

\newcommand{\tSdt}{\tilde{\sigma}_{\Dt}}
\newcommand{\tSdtm}{\tilde{\sigma}^{\Dt}_{\rm main}}
\newcommand{\tSdtt}{\tilde{\sigma}^{\Dt}_{\rm tail}}

\newcommand{\Sdtm}{\sigma^{\Dt}_{\rm main}}
\newcommand{\Mdtm}{\mu^{\Dt}_{\rm main}}
\newcommand{\Sdtt}{\sigma^{\Dt}_{\rm tail}}
\newcommand{\Mdtt}{\mu^{\Dt}_{\rm tail}}

\newcommand{\tScp}{\tilde{\sigma}_{CP}}
\newcommand{\tStag}{\tilde{\sigma}_{\rm tag}}
\newcommand{\tSdz}{\tilde{\sigma}_{\Dz}}
\newcommand{\tSdzm}{\tilde{\sigma}^{\Dz}_{\rm main}}
\newcommand{\tSdzt}{\tilde{\sigma}^{\Dz}_{\rm tail}}

\newcommand{\Sdzm}{\sigma^{\Dz}_{\rm main}}
\newcommand{\Mdzm}{\mu^{\Dz}_{\rm main}}
\newcommand{\Sdzt}{\sigma^{\Dz}_{\rm tail}}
\newcommand{\Mdzt}{\mu^{\Dz}_{\rm tail}}

\newcommand{\srec}{s_{\rm rec}}
\newcommand{\sasc}{s_{\rm asc}}

\newcommand{\sk}{\sigma_{\rm k}}
\newcommand{\snpm}{s_{\rm main}^{\rm NP}}
\newcommand{\snpt}{s_{\rm tail}^{\rm NP}}
\newcommand{\smain}{s_{\rm main}}
\newcommand{\stail}{s_{\rm tail}}
\newcommand{\ftail}{f_{\rm tail}}

\newcommand{\bra}[1]{\langle#1|}
\newcommand{\ket}[1]{|#1\rangle}
\newcommand{\braket}[2]{\langle#1|#2\rangle}
\newcommand{\rexp}[1]{{\rm e}^{#1}}
\newcommand{\ri}{{\rm i}}
\newcommand{\lcp}{{\lambda_{CP}}}
\newcommand{\dedx}{{\rm d}E/{\rm d}x}
\newcommand{\sinbb}{{\sin2\phi_1}}

\newcommand{\bcdot}{\!\cdot\!}

\def\dzp{\delz^\prime}
\def\fdelbg{f_\delta^{\rm bkg}}
\def\mudelbg{\mu_\delta^{\rm bkg}}
\def\mutaubg{\mu_\tau^{\rm bkg}}

\def\pol{p_{\rm ol}}
\def\fol{f_{\rm ol}}
\def\sigol{\sigma_{\rm ol}}

\newcommand{\ra}{\rightarrow}
\newcommand{\myindent}{\hspace*{2cm}}  
\newcommand{\fCP}{f_{CP}}
\newcommand{\ftag}{f_{\rm tag}}
\newcommand{\zCP}{z_{CP}}
\newcommand{\tCP}{t_{CP}}
\newcommand{\ttag}{t_{\rm tag}}
\newcommand{\cala}{{\cal A}}
\newcommand{\cals}{{\cal S}}
\newcommand{\dm}{\Delta m_d}
\newcommand{\dmd}{\dm}
\def\taubz{\tau_\bz}
\def\ks{{K_S^0}}
%
%

\title{\boldmath Study of Time-Dependent {\boldmath $CP$}-Violating Asymmetries 
       \\in $b\to s\overline{q}q$ Decays}

\date{\today}

\affiliation{Budker Institute of Nuclear Physics, Novosibirsk}
\affiliation{Chiba University, Chiba}
\affiliation{Chuo University, Tokyo}
\affiliation{University of Cincinnati, Cincinnati, Ohio 45221}
\affiliation{University of Frankfurt, Frankfurt}
\affiliation{Gyeongsang National University, Chinju}
\affiliation{University of Hawaii, Honolulu, Hawaii 96822}
\affiliation{High Energy Accelerator Research Organization (KEK), Tsukuba}
\affiliation{Hiroshima Institute of Technology, Hiroshima}
\affiliation{Institute of High Energy Physics, Chinese Academy of Sciences, Beijing}
\affiliation{Institute of High Energy Physics, Vienna}
\affiliation{Institute for Theoretical and Experimental Physics, Moscow}
\affiliation{J. Stefan Institute, Ljubljana}
\affiliation{Kanagawa University, Yokohama}
\affiliation{Korea University, Seoul}
\affiliation{Kyoto University, Kyoto}
\affiliation{Kyungpook National University, Taegu}
\affiliation{Institut de Physique des Hautes \'Energies, Universit\'e de Lausanne, Lausanne}
\affiliation{University of Ljubljana, Ljubljana}
\affiliation{University of Maribor, Maribor}
\affiliation{University of Melbourne, Victoria}
\affiliation{Nagoya University, Nagoya}
\affiliation{Nara Women's University, Nara}
\affiliation{National Lien-Ho Institute of Technology, Miao Li}
\affiliation{National Taiwan University, Taipei}
\affiliation{H. Niewodniczanski Institute of Nuclear Physics, Krakow}
\affiliation{Nihon Dental College, Niigata}
\affiliation{Niigata University, Niigata}
\affiliation{Osaka City University, Osaka}
\affiliation{Osaka University, Osaka}
\affiliation{Panjab University, Chandigarh}
\affiliation{Peking University, Beijing}
\affiliation{Princeton University, Princeton, New Jersey 08545}
\affiliation{RIKEN BNL Research Center, Upton, New York 11973}
\affiliation{Saga University, Saga}
\affiliation{University of Science and Technology of China, Hefei}
\affiliation{Seoul National University, Seoul}
\affiliation{Sungkyunkwan University, Suwon}
\affiliation{University of Sydney, Sydney NSW}
\affiliation{Toho University, Funabashi}
\affiliation{Tohoku Gakuin University, Tagajo}
\affiliation{Tohoku University, Sendai}
\affiliation{University of Tokyo, Tokyo}
\affiliation{Tokyo Institute of Technology, Tokyo}
\affiliation{Tokyo Metropolitan University, Tokyo}
\affiliation{Tokyo University of Agriculture and Technology, Tokyo}
\affiliation{Toyama National College of Maritime Technology, Toyama}
\affiliation{University of Tsukuba, Tsukuba}
\affiliation{Virginia Polytechnic Institute and State University, Blacksburg, Virginia 24061}
\affiliation{Yokkaichi University, Yokkaichi}
\affiliation{Yonsei University, Seoul}
  \author{K.~Abe}\affiliation{High Energy Accelerator Research Organization (KEK), Tsukuba} 
  \author{T.~Abe}\affiliation{Tohoku University, Sendai} 
  \author{I.~Adachi}\affiliation{High Energy Accelerator Research Organization (KEK), Tsukuba} 
  \author{H.~Aihara}\affiliation{University of Tokyo, Tokyo} 
  \author{K.~Akai}\affiliation{High Energy Accelerator Research Organization (KEK), Tsukuba} 
  \author{M.~Akatsu}\affiliation{Nagoya University, Nagoya} 
  \author{M.~Akemoto}\affiliation{High Energy Accelerator Research Organization (KEK), Tsukuba} 
  \author{Y.~Asano}\affiliation{University of Tsukuba, Tsukuba} 
  \author{T.~Aso}\affiliation{Toyama National College of Maritime Technology, Toyama} 
  \author{V.~Aulchenko}\affiliation{Budker Institute of Nuclear Physics, Novosibirsk} 
% ...
\collaboration{The Belle Collaboration}

\begin{abstract}
We present a measurement of $CP$-violation parameters 
in the $b \to s\overline{q}q$ penguin transitions ($q=s,u,d$)
based on a 78~fb$^{-1}$ data sample collected at the 
$\Upsilon(4S)$ resonance with
the Belle detector at the KEKB energy-asymmetric $e^+e^-$ collider.
One neutral $B$ meson is reconstructed in the
$\phi\ks$, $K^+K^-\ks$, or $\eta'\ks$ decay channel, and
the flavor of the accompanying $B$ meson is identified from 
its decay products.
$CP$ violation parameters for each of the three
modes are obtained from the asymmetries in the distributions of
the proper-time intervals between the two $B$ decays.
\end{abstract}


\maketitle

In the standard model (SM), $CP$ violation arises in weak interactions from an irreducible complex phase
in the Kobayashi-Maskawa (KM) quark-mixing matrix \cite{bib:KM}. In
particular, the SM predicts $CP$-violating asymmetries in the time-dependent rates for $\bz$ and
$\bzb$ decays to a common $CP$ eigenstate $\fCP$~\cite{bib:sanda}. Recent measurements of the $CP$-violating
parameter $\sin2\phi_1$ by the Belle~\cite{bib:CP1_Belle,bib:Belle_sin2phi1_78fb-1}
and BaBar~\cite{bib:CP1_BaBar} collaborations established $CP$ violation
in neutral $B$ meson decays mediated by the $b \to c\overline{c}s$ tree transition~\cite{bib:CC}
at a level consistent with KM expectations.

Despite this success,  many tests remain before one can conclude that the KM model 
provides a complete description.  
For example, the charmless decays $\bz\to \phi\ks$, $\bz\to K^+K^-\ks$, and
$\bz\to \eta'\ks$, which are mediated by the $b\to s\overline{s}s$ transition 
($\bz\to \eta'\ks$ also receives $b\to s\overline{d}d$ and $b\to s\overline{u}u$ penguin contributions) 
are potentially sensitive to new $CP$-violating phases from physics beyond the SM~\cite{bib:lucy}.
SM contributions from the $b\to u\overline{u}s$ tree diagram
are expected to be highly suppressed~\cite{bib:tree-penguin,bib:Garmash}.
Thus, the SM predicts that $CP$ violation measurements in these charmless modes 
should yield $\sin 2\phi_1$ to a good approximation.
Consequently, a significant deviation in the time-dependent $CP$ asymmetry in these modes from what is
observed 
in $b \to c\overline{c}s$ decays would be evidence of a $CP$-violating phase not expected in the KM model.

In the decay chain $\Upsilon(4S)\to \bz\bzb \to f_{CP}f_{\rm tag}$,
where one of the $B$ mesons decays at time $t_{CP}$ to a final state $f_{CP}$ 
and the other decays at time $t_{\rm tag}$ to a final state  
$f_{\rm tag}$ that distinguishes between $B^0$ and $\bzb$, 
the decay rate has a time dependence
given by~\cite{bib:CPVrev}
\begin{equation}
\label{eq:psig}
{\cal P}(\Delta{t}) = 
\frac{e^{-|\Delta{t}|/{\taubz}}}{4{\taubz}}
\biggl\{1 + q\cdot 
\Bigl[ \cals\sin(\dmd\Delta{t})
   + \cala\cos(\dmd\Delta{t})
\Bigr]
\biggr\},
\end{equation}
where $\taubz$ is the $B^0$ lifetime, $\dmd$ is the mass difference 
between the two $B^0$ mass
eigenstates, $\Delta{t}$ = $t_{CP}$ $-$ $t_{\rm tag}$, and
the $b$-flavor charge $q$ = +1 ($-1$) when the tagging $B$ meson
is a $B^0$ 
($\bzb$).
The $CP$-violating parameters $\cals$ and $\cala$  are given by
\begin{equation}
\cals \equiv \frac{2{\cal I}m(\lambda)}{|\lambda|^2+1}, \qquad
\cala \equiv \frac{|\lambda|^2-1}{|\lambda|^2+1},
\end{equation}
where $\lambda$ is a complex 
parameter that depends on both the $\bz\bzb$
mixing and on the amplitudes for $\bz$ and $\bzb$ decay to $\fCP$.
To a good approximation in the SM, 
$|\lambda|$ is equal to the absolute value
of the ratio of the $\bzb\to\fCP$ to $\bz\to\fCP$ decay amplitudes.
The SM predicts $\cals = -\xi_f\sin 2\phi_1$,  where $\xi_f = +1 (-1)$ 
corresponds to  $CP$-even (-odd) final states; and $\cala =0$ (or 
equivalently $|\lambda| = 1$) for both $b \to c\overline{c}s$ and 
$b \to s\overline{s}s$ transitions. 

In this paper, we report the first measurement of
$CP$ asymmetries in $\bz \to \phi \ks$ and $K^+K^-\ks$ decays,
and an improved measurement for  $\bz \to \eta'\ks$ 
decay~\cite{bib:Belle_etapks}
based on a $78~{\rm fb}^{-1}$ data sample,
which contains 85 million $B\overline{B}$ pairs. 
Data are collected  with the Belle detector at the KEKB energy-asymmetric 
$e^+e^-$ (3.5 on 8~GeV) collider~\cite{bib:KEKB}
operating at the $\Upsilon(4S)$ resonance.
At KEKB, the $\Upsilon(4S)$ is produced
with a Lorentz boost of $\beta\gamma=0.425$ nearly along
the electron beamline ($z$).
Since the $B^0$ and $\bzb$ mesons are approximately at 
rest in the $\Upsilon(4S)$ center-of-mass system (cms),
$\Delta t$ can be determined from the displacement in $z$ 
between the $f_{CP}$ and $f_{\rm tag}$ decay vertices:
$\Delta t \simeq (z_{CP} - z_{\rm tag})/\beta\gamma c
 \equiv \Delta z/\beta\gamma c$.

The Belle detector~\cite{bib:Belle} is a large-solid-angle spectrometer
that includes a three-layer silicon vertex detector (SVD),
a 50-layer central drift chamber (CDC),
an array of aerogel threshold Cherenkov counters (ACC),
time-of-flight (TOF) scintillation counters,
and an electromagnetic calorimeter comprised of CsI(Tl) crystals (ECL)
located inside a superconducting solenoid coil
that provides a 1.5~T magnetic field.
An iron flux-return located outside of the coil is instrumented
to detect $\kl$ mesons and to identify muons (KLM).

We reconstruct $\bz$ decays to
$\phi\ks$ and $\eta'\ks$ final states for $\xi_f=-1$, and
$\bz\to K^+K^-\ks$ decays
that are a mixture of $\xi_f=+1$ and $-1$.
$K^+K^-$ pairs that are consistent with $\phi \to K^+K^-$ decay are excluded
from the $\bz \to K^+K^-\ks$ sample.
     We find that the $K^+K^-\ks$ state is primarily $\xi_f=+1$; the $\xi_f=+1$ fraction is
  $1.04 \pm 0.19\mbox{(stat)}\pm0.06\mbox{(syst)}$~\cite{bib:Garmash}.
In the following determination of $\cals$ and $\cala$,
we fix $\xi_f=+1$.
The intermediate meson states are reconstructed from the following decay chains:
$\eta'\to\rho^0 (\to \pi^+\pi^-) \gamma$ or $\eta'\to\pi^+\pi^-\eta (\to \gamma\gamma)$,
$\ks \to \pi^+\pi^-$, and $\phi\to K^+K^-$.

Candidate $\ks \to \pi^+\pi^-$ decays are oppositely charged track pairs that
have an invariant mass within 15 (12) MeV/$c^2$ of the nominal $\ks$ mass
for the $\bz \to \phi\ks$ ($\bz \to K^+K^-\ks$) mode.
The displacement of the $\pi^+\pi^-$ vertex from the
nominal interaction point (IP) in the plane transverse to the positron
beam axis ($r$-$\phi$ plane) is required to be greater than 0.1 cm and less than 20 cm.
The direction of the combined pion-pair momentum in the $r$-$\phi$ plane is
required to be within 0.2 radians of the direction defined by the IP and the 
displaced vertex.
We select candidate $\phi \to K^+K^-$ decays requiring that
the $K^+K^-$ invariant mass is within 10 MeV/$c^2$ of the nominal $\phi$ meson mass,
the $\phi$ meson momentum in the cms exceeds 2.0 GeV/$c$, and
the $K^+K^-$ vertex is consistent with the IP.
Since the $\phi$ meson selection is effective in reducing background events,
we impose only minimal kaon-identification requirements.
For selection of non-resonant $K^+K^-\ks$ candidates, 
more stringent kaon-identification requirements, which retain 86\% of $K^\pm$ at 
a 7\% fake rate for $\pi^\pm$, are used.  In this case,
charged tracks that are positively identified as electrons or protons are
excluded.  In addition to the rejection of $\phi$ meson candidates, we
reject $K^+K^-$ pairs that are consistent with 
$D^0 \to K^+K^-$ or $\chi_{c0} \to K^+K^-$ decay.

For reconstructed $B\to\fCP$ candidates, we identify $B$ meson decays using the
energy difference $\dE\equiv E_B^{\rm cms}-E_{\rm beam}^{\rm cms}$ and
the beam-energy constrained mass $\mb\equiv\sqrt{(E_{\rm beam}^{\rm cms})^2-
(p_B^{\rm cms})^2}$, where $E_{\rm beam}^{\rm cms}$ is
the beam energy in the cms, and
$E_B^{\rm cms}$ and $p_B^{\rm cms}$ are the cms energy and momentum of the 
reconstructed $B$ candidate, respectively.
The $B$ meson signal region is defined as 
$5.27~{\rm GeV}/c^2 <\mb<5.29~{\rm GeV}/c^2$ and
$|\dE|<0.051$ GeV for $\bz \to \phi \ks$ or
$|\dE|<0.040$ GeV for $\bz \to K^+K^-\ks$.
In order to suppress background from the $e^+e^- \rightarrow q\overline{q}$
($q = u,~d,~s$) and $c\overline{c}$ continuum,
we form signal and background
likelihood functions, ${\cal L}_{\rm S}$ and ${\cal L}_{\rm BG}$, 
from a set of variables that characterize the event 
topology~\cite{bib:Garmash,bib:Belle_etapks}. 
We determine ${\cal L}_{\rm S}$ from Monte Carlo (MC)
and ${\cal L}_{\rm BG}$ from data, and impose
mode-dependent thresholds on 
the likelihood ratio ${\cal L}_{\rm S}/({\cal L}_{\rm S}+{\cal L}_{\rm BG})$.
The numbers of reconstructed candidates are 59 and 230 for $\bz\to\phi\ks$ and for $\bz\to K^+K^-\ks$,
respectively.

For $\bz \to \eta'\ks$ decay,
we use the same selection criteria as those used in our previously published  
analysis~\cite{bib:Belle_etapks} 
if both of the charged pions in the $\eta'\to\pi^+\pi^-\eta$ or
the $\eta'\to\rho^0\gamma$ decay have associated SVD hits.
We also reconstruct events where
only one of the charged pions has associated SVD hits.
In this case, the requirement on the impact parameter
is relaxed for the track without SVD hits, while 
a higher threshold is imposed on the likelihood ratio.
The number of reconstructed $\bz\to\eta'\ks$ candidates is 311.

Charged leptons, kaons, pions, and $\Lambda$ baryons
that are not associated with the reconstructed $\fCP$ decay
are used to identify the $b$-flavor of the accompanying $B$ meson, which decays into $\ftag$. 
Based on the measured properties of these tracks, two parameters,
$q$ and $r$, are assigned to each event.
The first, $q$, has the discrete value $+1$~($-1$)
when the tag-side $B$ meson is more likely to be a $\bz$~($\bzb$).
The parameter $r$ is an event-by-event MC-determined
flavor-tagging dilution factor that ranges
from $r=0$ for no flavor discrimination
to $r=1$ for an unambiguous flavor assignment.
It is used only to sort data into six intervals of $r$,
according to the estimated flavor purity.
The wrong-tag probabilities for each of these intervals, $w_l~(l=1,6)$,
which are used in the final fit, are determined directly from the data.
Samples of $\bz$ decays to exclusively reconstructed self-tagging channels
are utilized to obtain $w_l$ using
time-dependent $\bz$-$\bzb$ mixing:
$(N_{\rm OF}-N_{\rm SF})/(N_{\rm OF}+N_{\rm SF}) = (1-2w_l)\cos(\dm\Dt)$,
where $N_{\rm OF}$ and $N_{\rm SF}$ are the numbers of opposite
($\bz\bzb\to\bz\bzb$) and same ($\bz\bzb\to\bz\bz,\bzb\bzb$) flavor events.
The event fractions and wrong tag fractions for each $r$ interval
are described elsewhere~\cite{bib:Belle_sin2phi1_78fb-1}.

The decay vertices of $\bz$ mesons are reconstructed using
tracks that have enough SVD hits: i.e. both $z$ and $r$-$\phi$
hits in at least one SVD layer and at least one additional
layer with a $z$ hit, where the $r$-$\phi$ plane is
perpendicular to the $z$ axis.
Each vertex position is required to be consistent with
the IP profile, which is determined run-by-run and
smeared in the $r$-$\phi$ plane by 21~$\mu$m
to account for the $B$ meson decay length.
With these requirements, we are able to determine a vertex
even with a single track.
The vertex position for the $\fCP$ decay is reconstructed
using charged kaons for $\bz\to \phi\ks$ and $\bz \to K^+K^-\ks$ decays
and using charged pions from $\rho^0$ or $\eta'$ decays
for $\bz \to \eta'\ks$.
The algorithm for
the $\ftag$ vertex reconstruction is chosen to minimize
the effect of long-lived particles, secondary vertices from
charmed hadrons and a small fraction of poorly reconstructed 
tracks~\cite{bib:resol_nim}. 
From all the charged tracks with associated SVD hits
except those used for $\fCP$, we select tracks
with a position error in the $z$ direction of less than 500 $\mu$m, and with an impact parameter
with respect to the $\fCP$ vertex of less than 500 $\mu$m.
Track pairs with opposite charges 
are removed if they form a $\ks$ candidate with 
an invariant mass within $\pm15$~MeV/$c^2$ of the nominal
$\ks$ mass.
If the reduced $\chi^2$ associated with the $\ftag$ vertex exceeds 20, the
track making the largest $\chi^2$ contribution is removed and the vertex is refitted.
This procedure is repeated until an acceptable reduced $\chi^2$ is obtained.

\begin{figure}
\resizebox{0.29\textwidth}{!}{\includegraphics{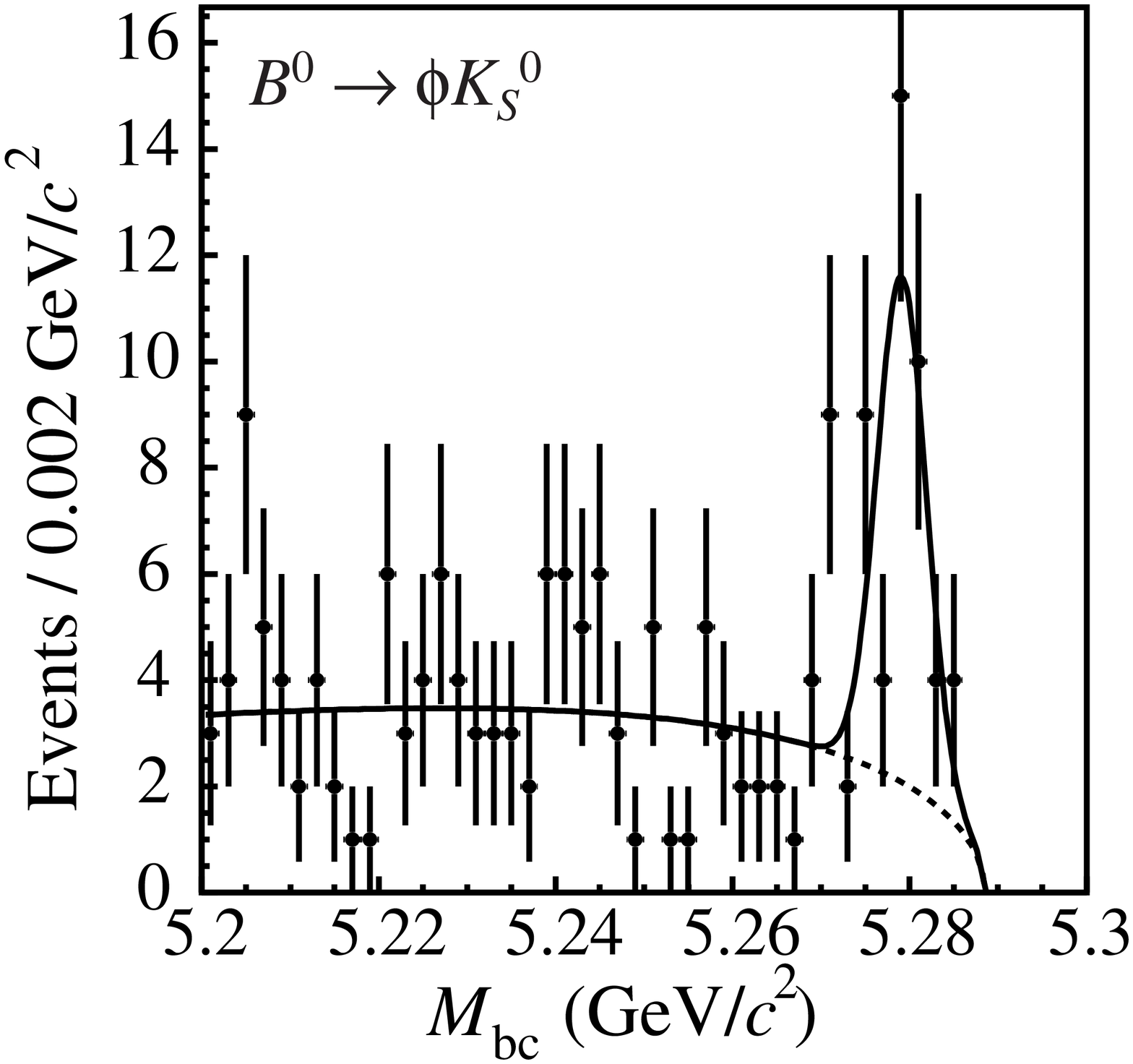}}
\hspace*{2mm}
\resizebox{0.29\textwidth}{!}{\includegraphics{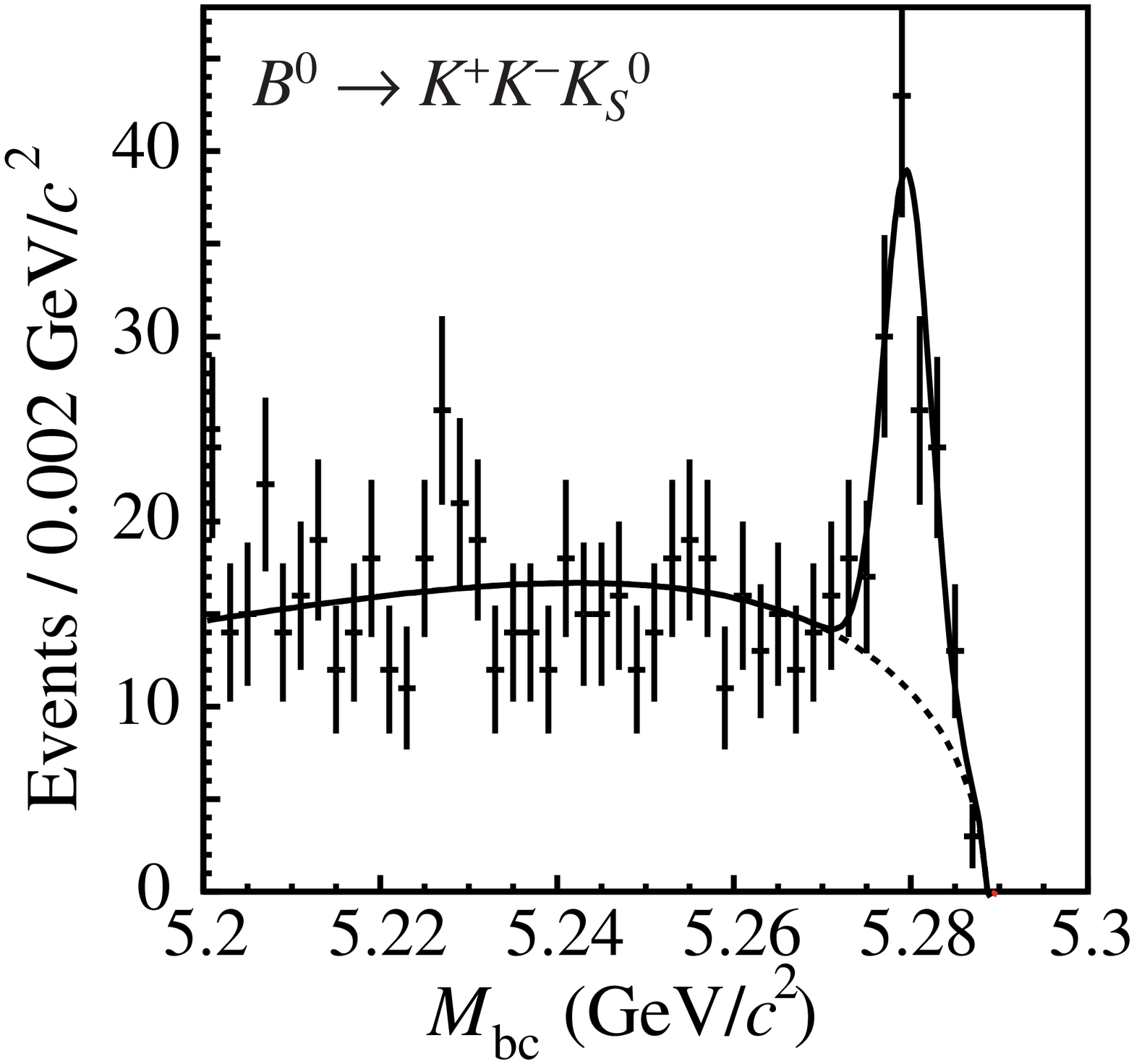}}
\hspace*{2mm}
\resizebox{0.29\textwidth}{!}{\includegraphics{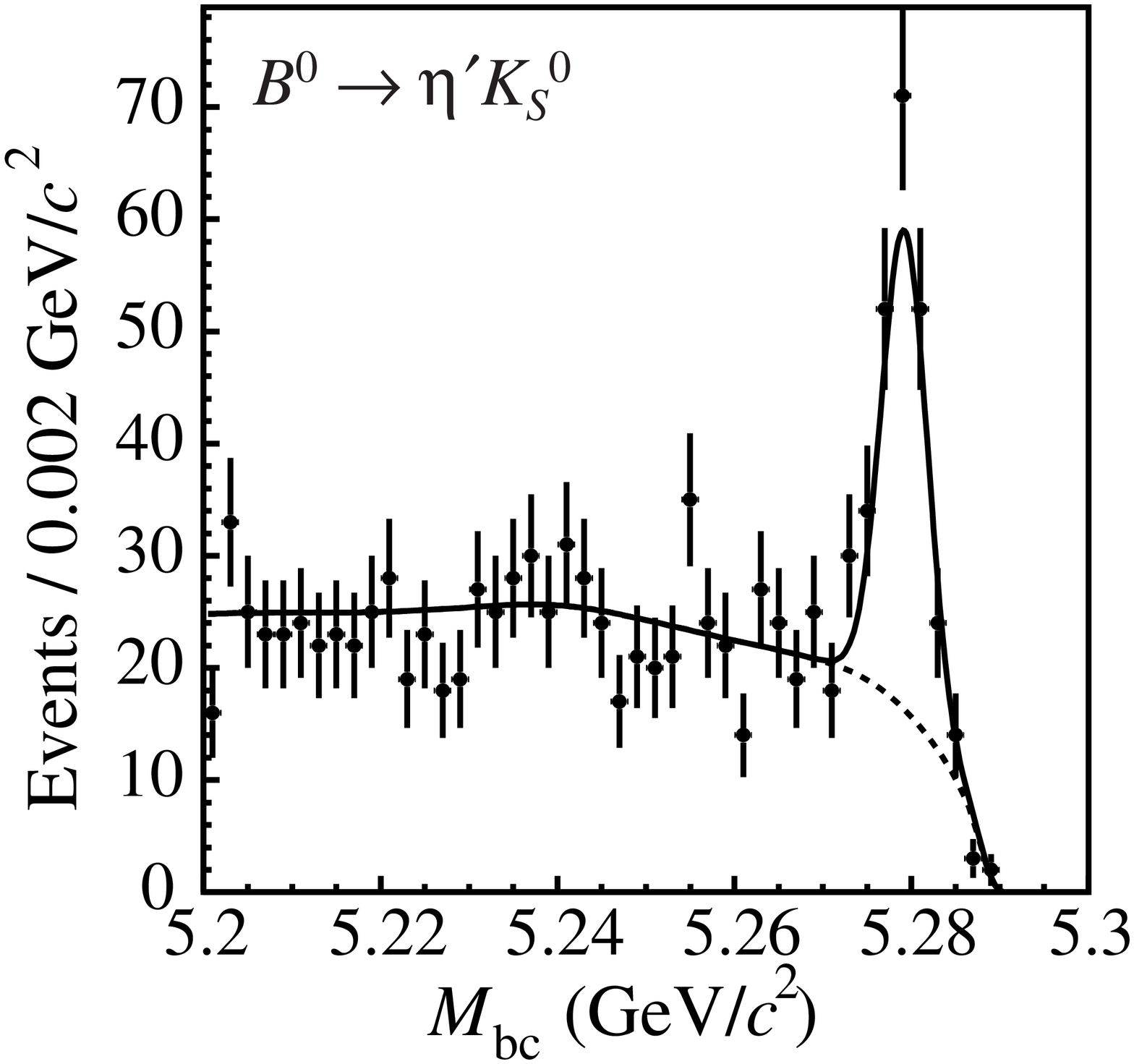}}
\caption{The beam-energy constrained mass distributions for
$\bz\to\phi\ks$ (left),
$\bz\to K^+K^-\ks$ (center), and
$\bz\to\eta'\ks$ (right)
within the $\dE$ signal region.
Solid curves show the fit to signal plus background distributions,
and dotted curves show the background contributions.
The background for $\bz\to\eta'\ks$ decay includes
an MC-estimated $B\overline{B}$ background component.
}\label{fig:mb}
\end{figure}

After flavor tagging and vertex reconstruction, we obtain the numbers of 
$\bz\to\fCP$ candidates, $N_{\rm ev}$,
listed in Table~\ref{tab:num}.
\begin{table}
\caption{
The numbers of reconstructed $\bz\to\fCP$ candidates
used for $\cals$ and $\cala$ determination, $N_{\rm ev}$, and the estimated
signal purity in the $\dE$-$\mb$ signal region for each $\fCP$ mode.}
\label{tab:num}
\begin{ruledtabular}
\begin{tabular}{lcrl}
\multicolumn{1}{c}{Mode} & $\xi_f$ & $N_{\rm ev}$ & Purity \\
\hline
$\phi\ks$   & $-1$                   &  53  & $0.67^{+0.07}_{-0.05}$\\
$K^+K^-\ks$ & $+1(100\%)$            & 191 & $0.50^{+0.04}_{-0.03}$\\
$\eta'\ks$  & $-1$                   & 299 & $0.49\pm0.05$\\
\end{tabular}
\end{ruledtabular}
\end{table}
Figure~\ref{fig:mb} shows the $\mb$ distributions for the reconstructed $B$ candidates that have $\dE$
values within the signal region.    
To assign an event-by-event signal probability for use in the maximum-likelihood fit of the
$CP$-violating parameters, we determine event distribution functions in the $\dE$-$\mb$ plane for
both signal and background.
The signal distribution is
modeled with a  single two-dimensional Gaussian, where the widths are allowed
to float in the fit to data.  
For the continuum background, we use a linear function
for $\dE$ and the ARGUS parameterization~\cite{bib:ARGUS} for $\mb$.
We use events outside the signal region 
as well as a large MC sample to study the background components.
The dominant background comes from continuum events.
In addition, according to MC simulation, there is  a non-negligible ($\sim 8\%$) contamination from $B\overline{B}$ background events
in $\bz\to\eta'\ks~(\eta'\to\rho^0\gamma)$. 
The contributions from $B\overline{B}$ events are smaller for other decay modes.
The contamination of $K^+ K^- \ks$ events in the $\phi\ks$ sample 
(and vice versa) is also small and is treated as a source of systematic uncertainty.
Finally, backgrounds from $B^0 \rightarrow f_0(980) \ks$ decay, which has the opposite 
$CP$ eigenvalue to $\phi\ks$, are found to be negligible.

\begin{figure}
\begin{center}
\resizebox{0.29\textwidth}{!}{\includegraphics{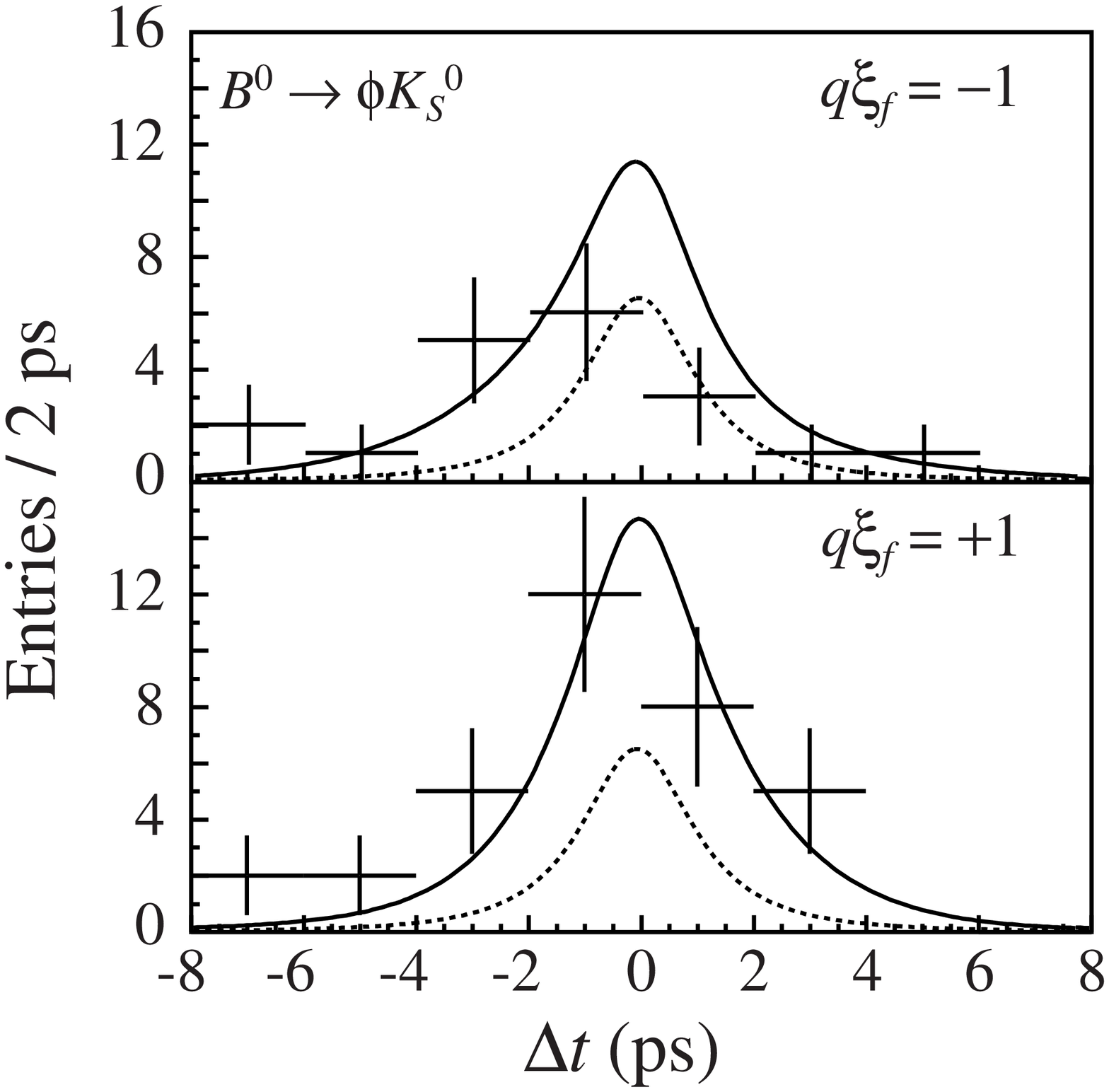}}
\hspace*{2mm}
\resizebox{0.29\textwidth}{!}{\includegraphics{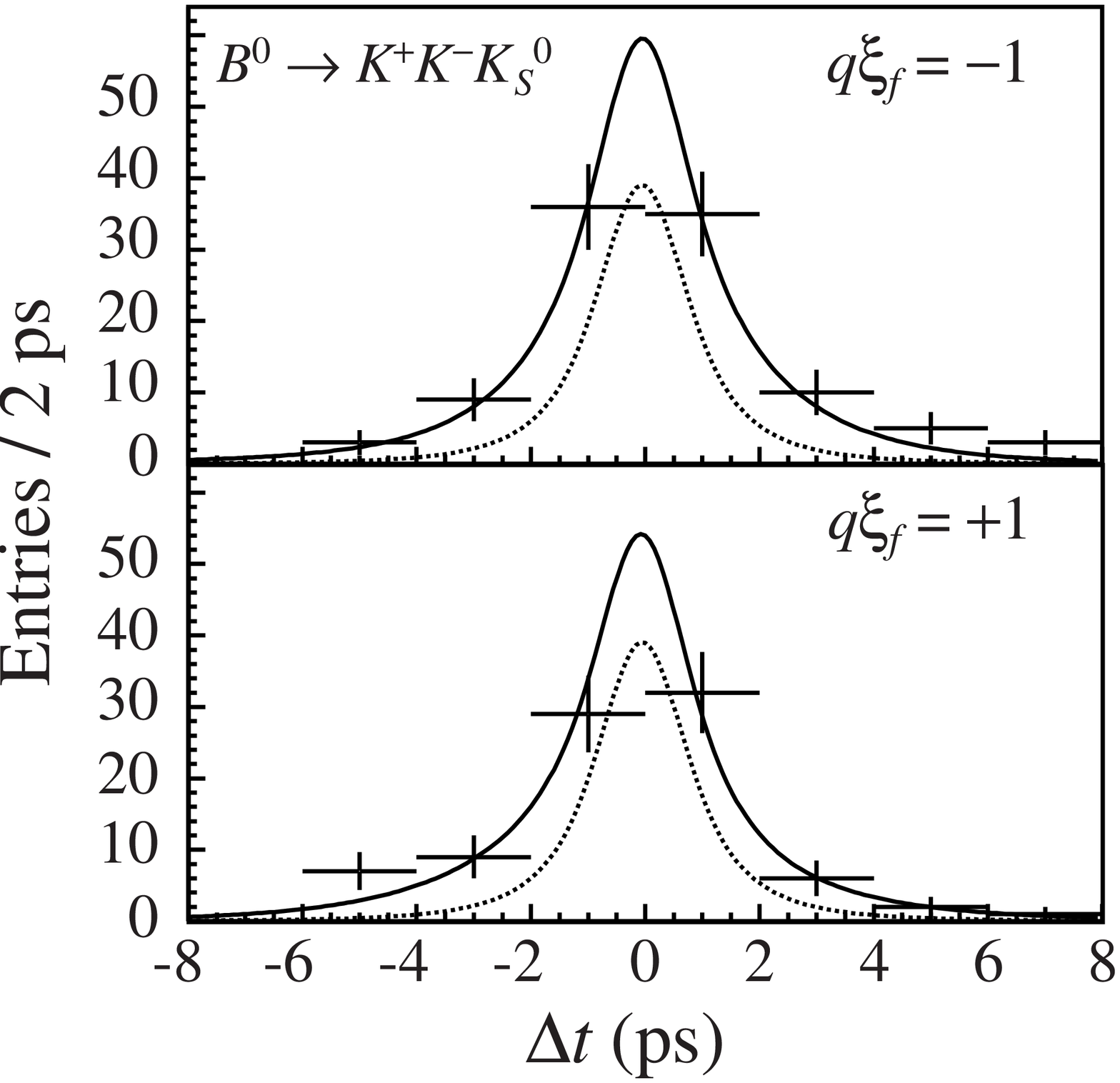}}
\hspace*{2mm}
\resizebox{0.29\textwidth}{!}{\includegraphics{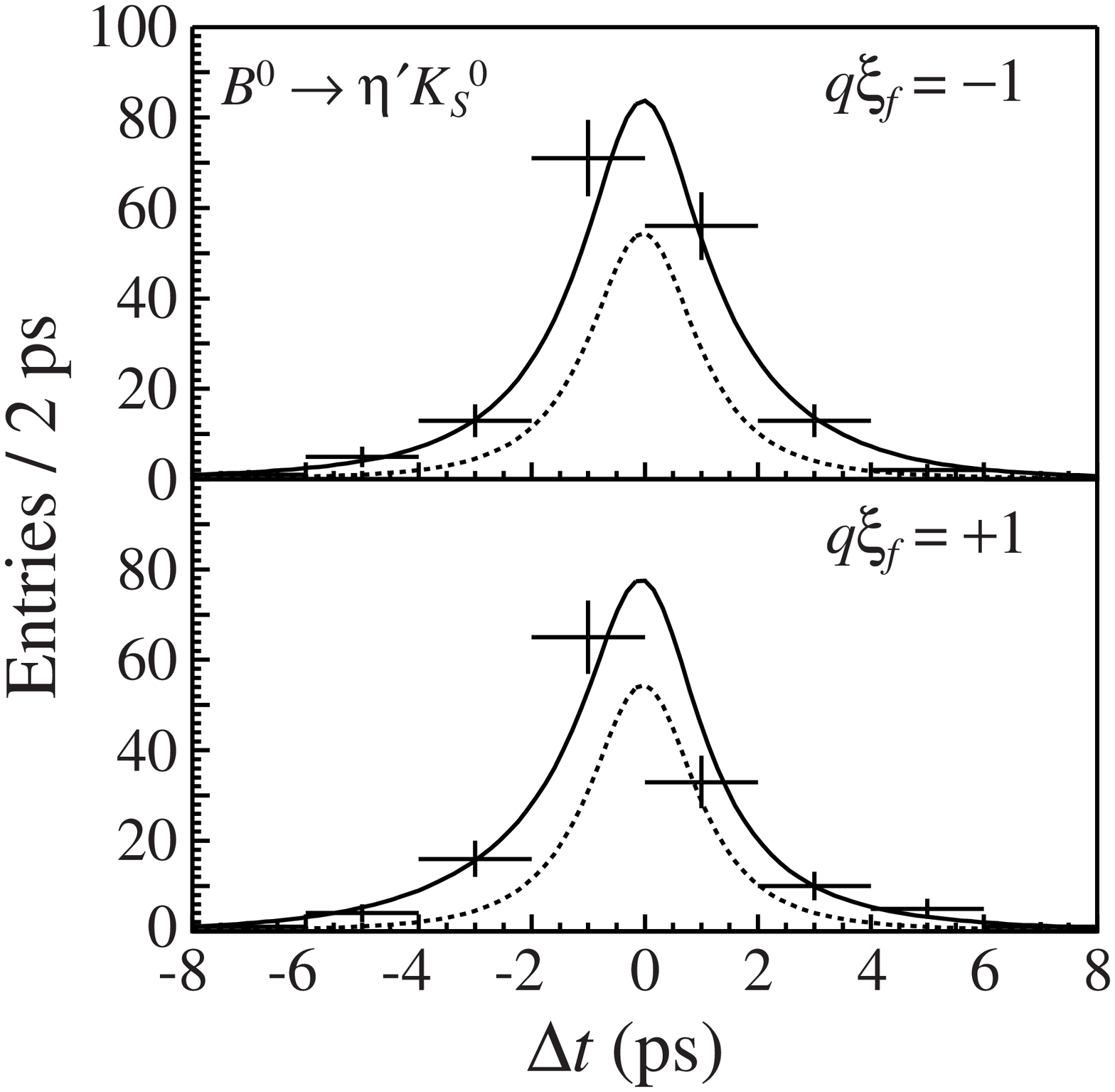}}
\end{center}
\caption{
The $\Dt$ distributions for 
$\bz\to\phi\ks$ (left), $\bz\to K^+K^-\ks$ (center),
and $\bz\to \eta'\ks$ (right) decays.  The upper and lower plots are for $q\xi_f=-1$ and
$q\xi_f=+1$ candidates, respectively.
The solid curves show the results of the global fits,
and dashed curves show the background distributions.
}
\label{fig:dt}
\end{figure}

\begin{figure}
\begin{center}
\begin{tabular}{ccccc}
\resizebox{!}{0.20\textwidth}{\includegraphics{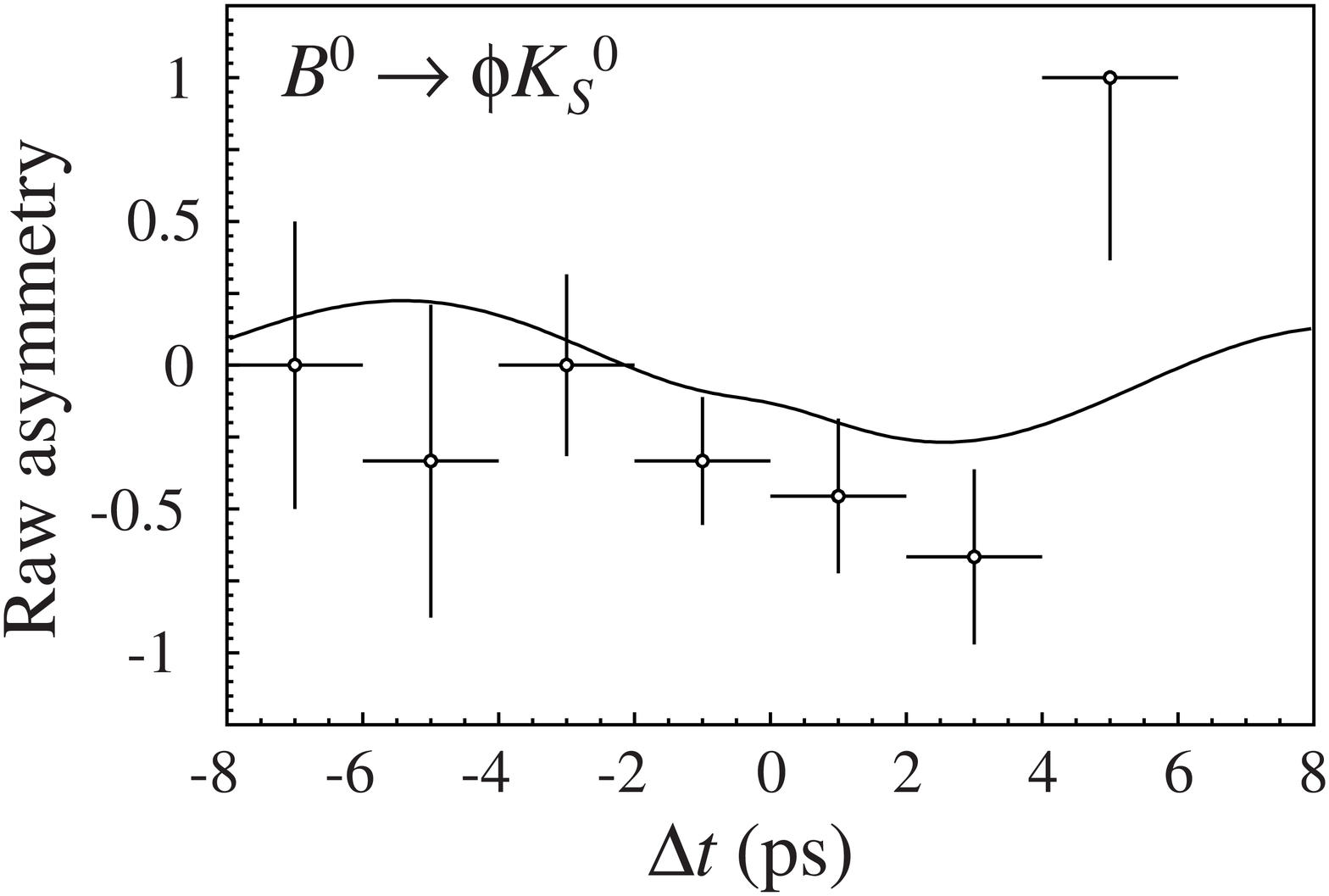}} &
\hspace*{.1cm} &
\resizebox{!}{0.20\textwidth}{\includegraphics{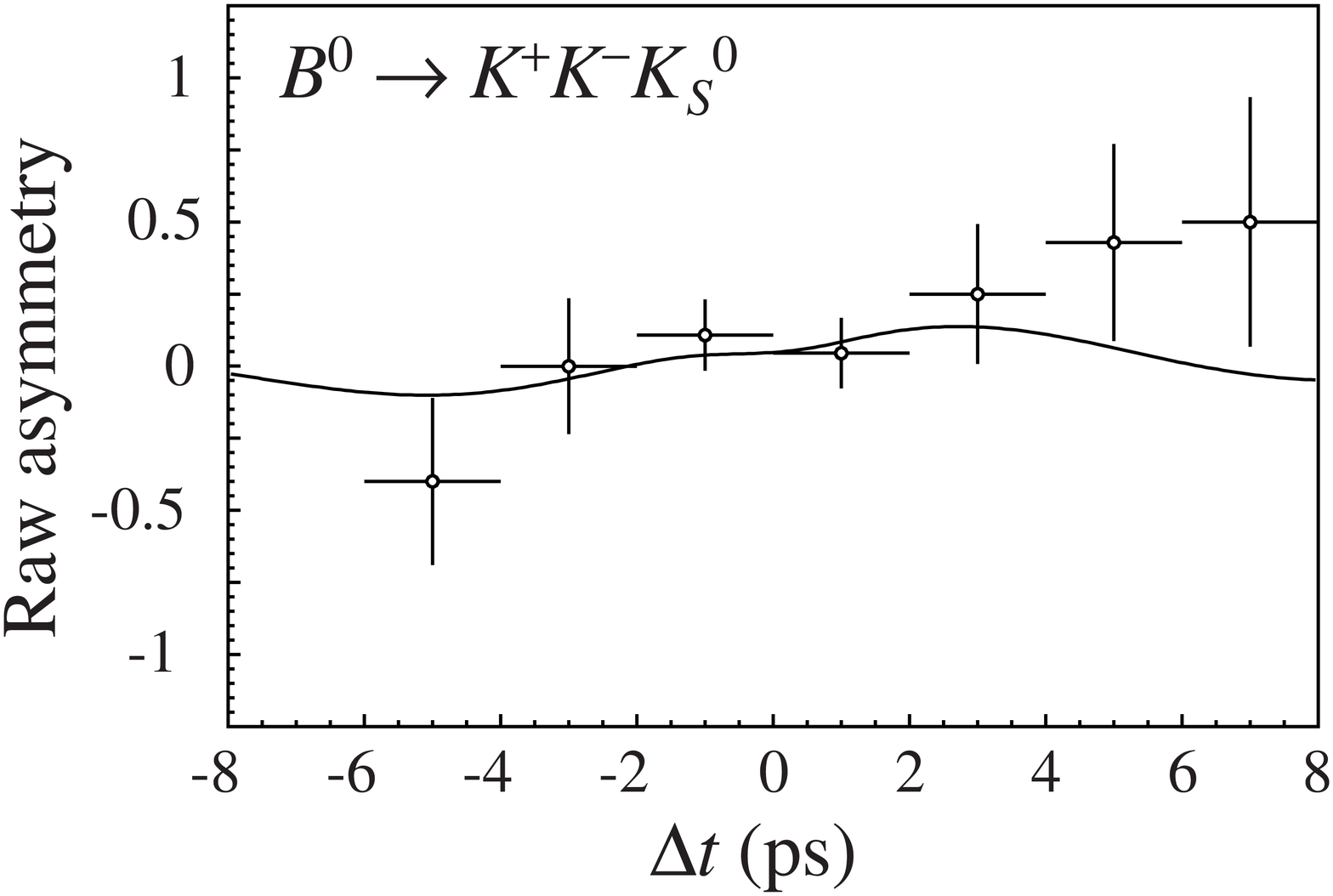}} &
\hspace*{.1cm} &
\resizebox{!}{0.20\textwidth}{\includegraphics{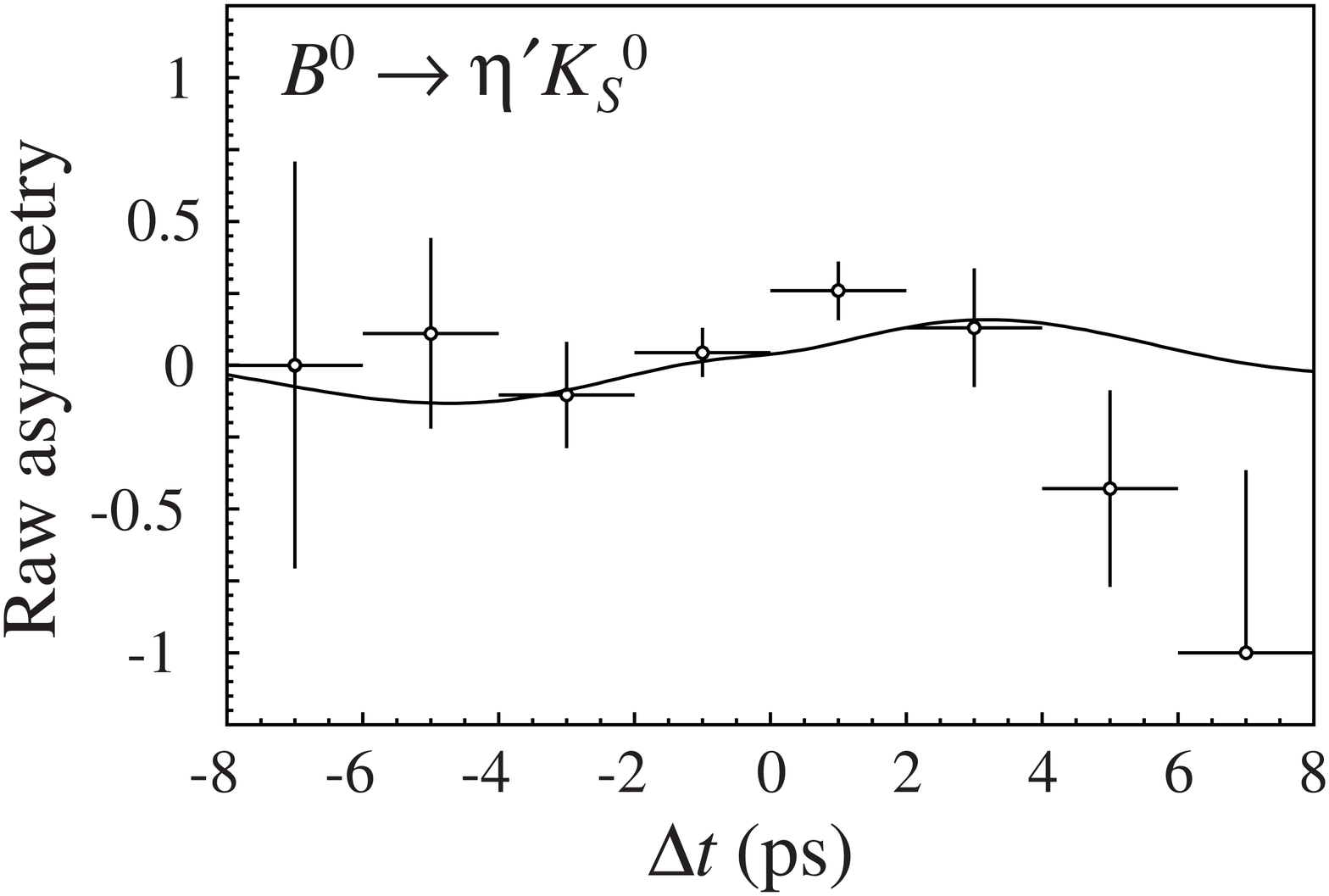}} 
\end{tabular}
\caption{The $\Dt$ asymmetry, $A$, in each bin for
$\bz\to\phi\ks$ (left),
$\bz\to K^+K^-\ks$ (center), and
$\bz\to \eta'\ks$ (right), respectively.
The curves show the results of the unbinned-maximum likelihood fit.}
\label{fig:asym}
\end{center}
\end{figure}

We determine $\cals$ and $\cala$ for each mode by performing an unbinned
maximum-likelihood fit to the observed $\Dt$ distribution.
The probability density function (PDF) expected for the signal
distribution is given by Eq.~(\ref{eq:psig}) with $q$ replaced by
$q(1-2w_l)$ to account for the effect of incorrect flavor assignment.
The distribution is
convolved with the
proper-time interval resolution function $R_{\rm sig}(\Dt)$,
which takes into account the finite vertex resolution. 
It is formed by convolving four components: the detector resolutions for $\zCP$ and
$\ztag$, the shift in the $\ztag$ vertex position 
due to secondary tracks originating from charmed particle decays, and 
the kinematic approximation that the $B$ mesons are at rest in the 
cms~\cite{bib:resol_nim}.
A small
component of broad outliers in the $\Dz$ distribution, caused by
mis-reconstruction, is represented by a Gaussian function $P_{\rm ol}(\Dt)$.
We determine twelve resolution parameters and the neutral- and charged-$B$ lifetimes simultaneously
from a fit to the $\Dt$ distributions of hadronic $B$ decays and
obtain an average $\Dt$ resolution of $\sim 1.43~$ps (rms).
We determine the following likelihood value for each
event:
\begin{eqnarray}
P_i(\Dt_i;\cals,\cala)
= (1-\fol)\int_{-\infty}^{\infty}\biggl[
\fsig{\cal P}_{\rm sig}(\Dt',q,w_l)R_{\rm sig}(\Dt_i-\Dt') & \nonumber \\
\quad +\; (1-\fsig){\cal P}_{\rm bkg}(\Dt')R_{\rm bkg}(\Dt_i-\Dt')\biggr]
d(\Dt') + \fol P_{\rm ol}(\Dt_i)
&
\end{eqnarray}
where $\fol$ is the outlier fraction and
$\fsig$ is the signal probability calculated as a function
of $\dE$ and $\mb$. 
${\cal P}_{\rm bkg}(\Dt)$ is a PDF for background events,
which dilutes the significance of $CP$ violation in Eq.\ (\ref{eq:psig}).
It is modeled as a sum of exponential and prompt components, and
is convolved with a sum of two Gaussians, $R_{\rm bkg}$, which represents a
resolution function for the background.
All parameters in ${\cal P}_{\rm bkg} (\Dt)$
and $R_{\rm bkg}$ are determined by the fit to the $\Dt$ distribution of a 
background-enhanced control sample~\cite{bib:BBbg}; i.e. events away from the $\dE$-$\mb$ signal region.
We fix the $\tau_\bz$ and $\dmd$ at
their world-average values~\cite{bib:PDG}.
The only free parameters in the final fit
are $\cals$ and $\cala$, which are determined by maximizing the
likelihood function
\begin{equation}
L = \prod_iP_i(\Dt_i;\cals,\cala)
\end{equation}
where the product is over all events.
Table \ref{tab:result} summarizes the results of the fit.
The table shows the values of $\cala$ and $-\xi_f\cals$, which, in 
the SM, is equal to $\sin 2\phi_1$. 
\begin{table}

\caption{Results of the fits to the $\Dt$ distributions.
The first errors are statistical and the second
errors are systematic.  The third error for the $K^+K^-\ks$ mode arise from
the uncertainty in the fraction of the $CP$-odd component.}

\label{tab:result}
\begin{ruledtabular}
\begin{tabular}{cll}
Mode &  \multicolumn{1}{c}{$-\xi_f\cals$ ($= \sinbb$ in the SM)} & 
\multicolumn{1}{c}{$\cala$ (= 0 in the SM)} \\
\hline
$\phi\ks$   & $-0.73\pm0.64\pm0.22$          & $-0.56\pm0.41\pm0.16$ \\
$K^+K^-\ks$ & $+0.49\pm0.43\pm0.11^{+0.33}_{-0.00}$     & $-0.40\pm0.33\pm0.10^{+0.00}_{-0.26}$ \\
$\eta'\ks$  & $+0.71\pm0.37^{+0.05}_{-0.06}$ & $+0.26\pm0.22\pm0.03$ \\
\end{tabular}
\end{ruledtabular}
\end{table}
The first errors are statistical and the second
errors are systematic.  The third error for the $K^+K^-\ks$ mode arises from
the uncertainty in the fraction of the $CP$-odd component~\cite{bib:Garmash}.
Figure~\ref{fig:dt} shows the observed $\Dt$ distribution 
for $q\xi_f=-1$ (upper figure) and $q\xi_f=+1$ (lower figure) event samples
for each $\fCP$ mode.
Figure~\ref{fig:asym} shows the raw asymmetry in each $\Dt$ bin without background subtraction,
which is defined by
\begin{equation}
A\equiv\frac{N_{q\xi_f=-1}-N_{q\xi_f=+1}}{N_{q\xi_f=-1}+N_{q\xi_f=+1}},
\end{equation}
where $N_{q\xi_f=+1(-1)}$ is the number of observed candidates with $q\xi_f=+1(-1)$.
The curves show the results of the unbinned-maximum likelihood fit to the asymmetry distribution,
$-\xi_f{\cal S}\sin(\Delta m_d \Dt) - \xi_f{\cal A}\cos(\Delta m_d\Dt)$.
These are the first measurements of the $CP$ violation parameters for
$\bz\to\phi\ks$ and $\bz\to K^+K^-\ks$ decays.  The result for 
$\eta'\ks$ supersedes the previous result~\cite{bib:Belle_etapks}. 
We obtain values consistent with
the present world average of $\sin 2\phi_1 = +0.734\pm 0.054$~\cite{bib:Nir02}
in  $\bz \to K^+K^-\ks$ and $\eta'\ks$ decays, while a 2.1$\sigma$ deviation 
is observed in $\bz \to \phi \ks$ decay.

Fits to the same samples with the direct $CP$ violation parameter $\cala$ fixed at zero yield
$-\xi_f\cals = -0.83\pm0.72$(stat) for $\bz\to\phi\ks$,
$-\xi_f\cals = +0.59\pm0.47$(stat) for $\bz\to  K^+K^-\ks$, and
$-\xi_f\cals =  +0.77\pm0.38$(stat) for $\bz\to\eta' \ks$.
As a consistency check for the $\cals$ term, we select the charged 
$B$ meson decays $B^+ \to \phi K^+$ and $B^+ \to \eta' K^+$
and apply the same fit procedure.
We obtain 
$\cals = 0.05\pm0.32$(stat), $\cala = 0.29 \pm 0.24$(stat) 
for $B^+ \to \phi K^+$ decay and
$\cals = -0.03\pm0.20$(stat), $\cala = 0.05 \pm 0.13$(stat) 
for $B^+ \to \eta' K^+$ decay.
Both results on the $\cals$ term are consistent with no $CP$ asymmetry, as expected. 

The largest source of systematic error for the $\bz \to \phi\ks$ mode
is the uncertainty in
the signal fraction and the background $\Dt$ shape ($\pm0.17$ for $\cals$ and $\pm0.14$
for $\cala$ in total) determined from the events in the sideband regions
in the $\dE$-$\mb$ plane.
Other significant contributions come from uncertainties in
the vertex reconstruction, the resolution function parameters,
wrong tag fractions, $\taubz$, and $\dmd$.
We add each contribution in quadrature to obtain the total systematic
uncertainty.
Systematic uncertainties from these sources
are also examined for the other modes. 
We find that 
the largest uncertainties arise
from the vertex reconstruction
($\pm0.09$ for $\cals$ and $\pm0.08$ for $\cala$)
for the $\bz \to K^+K^-\ks$ mode, and 
from the resolution function parameters ($^{+0.03}_{-0.04}$ for $\cals$)
and the signal fraction ($\pm0.02$ for $\cala$)
for the $\bz\to \eta' \ks$ mode.

In summary, we have performed the first measurement of 
$CP$ violation parameters in the $\bz \to \phi \ks$ and $K^+K^-\ks$ decays.
We also provide an improved measurement for $\eta'\ks$ decay.
These modes are dominated by the $b \to s\overline{s}s$ transition and are
sensitive to possible new $CP$-violating phases.
Our results for $\bz\to\eta'\ks$ and $K^+K^-\ks$ are consistent with those obtained for
$\bz \to J/\psi \ks$ and other decays governed by the $b \to c\overline{c}s$ transition.
A $2.1\sigma$ deviation is observed for $\bz\to\phi\ks$.

\begin{acknowledgments}
We wish to thank the KEKB accelerator group for the excellent
operation of the KEKB accelerator.
We acknowledge support from the Ministry of Education,
Culture, Sports, Science, and Technology of Japan
and the Japan Society for the Promotion of Science;
the Australian Research Council
and the Australian Department of Industry, Science and Resources;
the National Science Foundation of China under contract No.~10175071;
the Department of Science and Technology of India;
the BK21 program of the Ministry of Education of Korea
and the CHEP SRC program of the Korea Science and Engineering Foundation;
the Polish State Committee for Scientific Research
under contract No.~2P03B 17017;
the Ministry of Science and Technology of the Russian Federation;
the Ministry of Education, Science and Sport of the Republic of Slovenia;
the National Science Council and the Ministry of Education of Taiwan;
and the U.S.\ Department of Energy.
\end{acknowledgments}


\begin{thebibliography}{999}
\bibitem{bib:CC}
Throughout this paper,
the inclusion of the charge conjugate mode decay is implied
unless otherwise stated.
\bibitem{bib:BBbg}
Parameters in ${\cal P}_{\rm bkg}(\Dt)$ and $R_{\rm bkg}$
for $B\overline{B}$ background events in $\bz\to\eta'\ks$ decay
are determined from MC simulation.
\end{thebibliography}
\end{document}